\documentclass[aip,reprint]{revtex4-1}

\usepackage{graphicx}
\usepackage{amsmath,bm}
\usepackage{mathtools}
\usepackage{hyperref}
\usepackage{gensymb}
\usepackage{siunitx}
\usepackage{xcolor}
\usepackage[capitalize]{cleveref}

\newcommand*\diff{\mathop{}\!\mathrm{d}}

\begin{document}



\title{Basic requirements for potential differences across solid--fluid interfaces}




\author{David Fertig}
 \affiliation{Institute of Physics, Norwegian University of Life Sciences, \AA s, Norway}

 \author{Adrian L. Usler}
 \affiliation{Institute of Physics, Norwegian University of Life Sciences, \AA s, Norway}

\author{Mathijs Janssen}
\email{mathijs.a.janssen@nmbu.no}
 \affiliation{Institute of Physics, Norwegian University of Life Sciences, \AA s, Norway}

\date{\today}

\begin{abstract}
At model water--vapor and water--solid interfaces, molecular ordering leads to charge oscillations and, thereby, to a spatially varying electrostatic potential.
Atomistic simulations indicate that such ordering leads to an electric potential difference $\chi$, the surface potential, of about \SI{-0.5}{\volt} across the first few molecular layers. 
Here, we calculate surface potentials at interfaces between a simple model fluids and a solid, with Molecular Dynamics simulations. 
The fluids are made up of either diatomic, dipolar molecules or a single Lennard-Jones particle with a dipole moment.
All fluids show some structuring near the interface, but charge oscillations and a non-zero surface potential are present only for asymmetric molecules (unequal diameters of the atoms) or molecules with an off-center dipole.
We condense this finding into the criterion that the geometric and dipolar centers of a molecule must differ for the fluid to exhibit a surface potential.
Remarkably, while the solid--fluid interaction strength strongly affects the magnitude of charge oscillations, it hardly affects the potential drop $\chi$.
Further, our results demonstrate that changing the diameter of the smaller atom can flip the sign of the surface potential, thus highlighting the importance of steric effects.
\end{abstract}

\maketitle

\section{Introduction}
The interface between a solid and a fluid is the site of various phenomena, from adsorption and wetting to electrochemical reactions \cite{israelachvili2010intermolecular,Bard_2001_book}.
These physicochemical processes are governed by short-ranged molecular interactions, which are difficult to probe experimentally.
On the other hand, the structure of the solid--fluid interface has been studied extensively through molecular simulations~\cite{snook_jcp_1978,toxvaerd_fscs_1981,calleja_mp_1991,siepmann_jcp_1995,boda_jcp_1998,segura_jcp_1998,kislenko_pccp_2009,willard_fd_2009,becker_lm_2014,jiang_nr_2016,berg_jctc_2017,geada_nc_2018,scalfi_arpc_2021,liu_prl_2023,nickel_prl_2024,fertig_jpcc_2024,hisama_jcc_2024,joll_jpcl_2025,ahrensivers_jpc_2025}, integral equation theories~\cite{snook_jcp_1978,calleja_mp_1991,vossen_jcp_1994,segura_jcp_1998,xu_jpcb_2012}, and classical and density functional theories~\cite{segura_jcp_1998,carrasco_nm_2012,evans_jcp_2017,sauer_iecr_2017,clabaut_jctc_2020,bramley_jctc_2022,borgis_cs_2023}.
Quantum chemical density functional theory, for instance, gives access to interfacial energies, surface coverage, and solvent orientation at solid--liquid interfaces~\cite{carrasco_nm_2012,clabaut_jctc_2020}; calculating a fluid's interfacial structure over several molecular distances are not feasible.
Such larger length scales can be covered, still with atomistic detail, by molecular simulations.
Most classical Molecular Dynamics (MD) simulations of the solid--fluid interface employed two-body interactions~\cite{siepmann_jcp_1995,kislenko_pccp_2009,willard_fd_2009,jiang_nr_2016,berg_jctc_2017,geada_nc_2018,scalfi_arpc_2021}; many-body interactions have been incorporated recently by neural network potentials~\cite{liu_prl_2023,hisama_jcc_2024,joll_jpcl_2025}.
Moreover, ab initio MD can describe the metal--fluid interface accurately, but its high computational cost limits this method to small systems~\cite{heenen_jcp_2020,le_jacsau_2021,partanen_pccp_2024,dominguez_jcp_2024,xu_jcp_2024}.

An advantage of classical force fields is that the interaction parameters in simulated \emph{model fluids} can be varied systematically.
Becker and coworkers~\cite{becker_lm_2014}, for instance, studied droplets of Lennard-Jones (LJ) truncated-and-shifted particles on surfaces; varying the wall--fluid interactions led to different contact angles.
Moreover, Toxvaerd simulated a solid--liquid interface, with LJ interactions among all particles, and found density oscillations in the fluid near the solid~\cite{toxvaerd_fscs_1981}. 
These density oscillations are more pronounced in liquids than in vapors, in which at most two adsorbed layers are typically visible~\cite{heier_jced_2018,fertig_jpcc_2024}.

Mass density oscillations near interfaces can go along with charge density oscillations, even for net neutral molecules.
In mixtures, or in pure fluids with molecules composed of dissimilar atoms, different interactions between a fluid's components and a surface can lead to preferential or competitive adsorption~\cite{israelachvili2010intermolecular}.
MD simulations of the vapor--liquid interface of water showed that local polarization gives rise to an oscillating charge density \cite{sokhan_mp_1997,chapman_pccp_2022}.
In turn, these charge densities cause a spatially varying electrostatic potential, and a potential drop $\chi$ of about \SI{-0.5}{\volt} across the interface.
Likewise, simulations of water near graphite electrodes found charge oscillations in the fluid, again leading to a surface potential $\chi\sim \SI{-0.5}{\volt}$ \cite{nickel_prl_2024}.
Both for these vapor--liquid and solid--liquid simulations, $\chi$ decreased with temperature by about $\diff\chi/\diff T=-\SI{1.25}{\milli\volt/\kelvin}$.
While mass density oscillations have been probed experimentally \cite{auer_jcp_2023}, neither charge density oscillations nor surface potentials have.

\begin{figure*}
    \centering
	\includegraphics[width =0.98\linewidth]{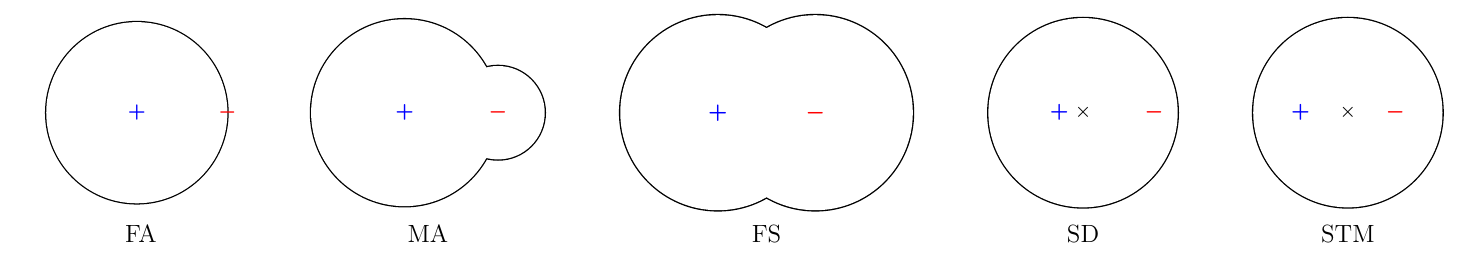}
    \caption{Molecular geometry and charge placement ($+$/$-$) with the respective labels employed in this study: fully asymmetric (FA), moderately asymmetric (MA), fully symmetric (FS), shifted dipole (SD), and Stockmayer-like (STM).}\label{fig:particles}
\end{figure*}

As Refs.~\cite{sokhan_mp_1997,chapman_pccp_2022,nickel_prl_2024,ahrensivers_jpc_2025} all dealt with water, the question arises what are the minimal requirements, in terms of geometry and charge distribution, for a fluid to exhibit charge density oscillations and a surface potential.
A recent article found that a model fluid of spherically symmetric molecules with off-center dipole moments displays a potential difference across the vapor--liquid interface \cite{varner_arxiv_2025}.
Here, we investigate the solid--fluid interface using model polar fluids and a model solid.
Instead of an LJ center and a point dipole, as studied by \cite{varner_arxiv_2025}, we use point charges to create a dipole.
Moreover, we consider five model fluids, including three fluids with diatomic molecules, with different atomic size ratios.\vfill

\pagebreak
\section{Method}
\subsection{Molecular model}

\begin{figure}
    \centering
	\includegraphics[width =0.6\linewidth]{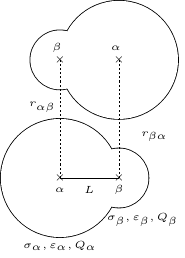}
    \caption{Interactions and molecular parameters, illustrated for the moderately asymmetric (MA) molecules.}\label{fig:setup}
\end{figure}

We consider five model fluids [\cref{fig:particles}] interacting through Lennard-Jones (LJ) and Coulomb pair potentials, 

\begin{subequations}
\begin{align}
&u(r_{ij},r_{kl},Q_k,Q_l)
  = u_{\mathrm{LJ}}(r_{ij})+ u_{\mathrm{C}}(r_{kl},Q_{k},Q_{l})
\intertext{where}
&u_{\mathrm{LJ}}(r_{ij})
  = \sum_{i,j}
     4\varepsilon_{ij}\left[
     \left(\frac{\sigma_{ij}}{r_{ij}}\right)^{12}
     -\left(\frac{\sigma_{ij}}{r_{ij}}\right)^{6}\right],
\intertext{and}
&u_{\mathrm{C}}(r_{kl},Q_{k},Q_{l})
  = \frac{1}{4\pi\epsilon_0}
     \sum_{k,l}
     \frac{Q_k Q_l}{r_{kl}},
\end{align}
\end{subequations}
where $u_{\mathrm{LJ}}$ is the LJ potential, $u_{\mathrm{C}}$ is the Coulomb potential, $r_{ij}$ is the distance between the centers of two atoms labeled by $i$ and $j$, $\sigma_{ij}$ and $\varepsilon_{ij}$ are the size and energy parameters of the LJ potential for interactions between atoms $i$ and~$j$, $\epsilon_0$ is the vacuum permittivity, and $Q_k$ is the charge placed at site $k$, see \cref{fig:setup}. 
The sums run over intermolecular interactions, not intramolecular interactions.

The first three fluids consist of diatomic, dipolar molecules.
For these fluids, the two atoms within a molecule, with centers $\alpha$ and $\beta$, have diameters $\sigma_\alpha$ and $\sigma_\beta$, with \mbox{$\sigma_\alpha\ge\sigma_\beta$}, see \cref{fig:setup}.
The two centers carry opposite point charges, \mbox{$Q_{\alpha}=-Q_{\beta}$}.
Not allowing vibrations, we keep the distance between the atomic centers fixed to a distance \mbox{$L/\sigma_\alpha=0.5$}.
The molecules of the three diatomic fluids differ in the size ratio of the atoms $\alpha$ and $\beta$, \mbox{$\sigma_{\beta}/\sigma_{\alpha}=0,0.5,\text{ and }1$} as well as in the ratio of the energy parameters \mbox{$\varepsilon_{\beta}/\varepsilon_{\alpha}=0,0.5,\text{ and }1$}.
Here, $\varepsilon_i$ is the LJ interaction parameter of atom $i$, with $i\in \{\alpha,\beta\}$.
We refer to these model fluids as FA (fully asymmetric, \mbox{$\sigma_{\beta}=0$}, $\varepsilon_{\beta}=0$), 
MA (moderately asymmetric, \mbox{$\sigma_{\beta}=0.5\sigma_{\alpha}$}, $\varepsilon_{\beta}=0.5\varepsilon_{\alpha}$), and FS (fully symmetric, \mbox{$\sigma_{\beta}=\sigma_{\alpha}$}, $\varepsilon_{\beta}=\varepsilon_{\alpha}$), see \cref{fig:particles}.
The FA molecule resembles model fluids with off-center dipole moments (such as SPC/E~\cite{berendsen_jpc_1987} and HCl~\cite{huang_aiche_2011}), MA resembles models of CO~\cite{stoll_jcp_2003} and NO~\cite{vanleeuwen_fpe_1994} with explicit charges instead of point dipoles, while FS does not resemble any real molecules.
When $\sigma_{\beta}=0$ and $\varepsilon_{\beta}=0$, we set $\sigma_{\alpha\beta}=0$ and $\varepsilon_{\alpha\beta}=0$.
For other interactions between different atoms in different fluid molecules, $\sigma_{ij}$ and $\varepsilon_{ij}$ are described by the Lorentz--Berthelot combination rules~\cite{lorentz_ap_1881,berthelot_crh_1898}, \mbox{$\sigma_{ij}=(\sigma_i+\sigma_j)/2$} and \mbox{$\varepsilon_{ij}=\sqrt{\varepsilon_i\varepsilon_j}$}.

For all three diatomic fluids, the masses of the two atoms have the same ratio, $m_{\beta}/m_{\alpha}=0.0625$.
Varying this ratio would affect the rotation of the model dipoles, not their equilibrium structure. 
We set the charges on the two sites to \mbox{$Q_{\alpha}=-Q_{\beta}=0.2Q_{\mathrm{LJ}}$}, where \mbox{$Q_{\mathrm{LJ}}=\sqrt{4\pi\epsilon_0\varepsilon\sigma}$} is the LJ unit of charge, resulting in a small dipole moment $\mu$ with \mbox{$\mu^2 / (\sigma Q_{\mathrm{LJ}})^2=0.01$}.
Throughout this study, we keep $\sigma_{\alpha}$ and $\varepsilon_{\alpha}$ fixed and, from here on, we define all quantities in terms of the LJ parameters of atom $\alpha$, and omit it in subscripts ($\sigma=\sigma_{\alpha}$, $\varepsilon=\varepsilon_{\alpha}$, $m=m_{\alpha}$, and $Q=Q_{\alpha}$).

The fourth and fifth model fluids are again dipolar, but monoatomic rather than diatomic. 
These fluids have one LJ center and two off-center charges, in two different configurations, see \cref{fig:particles}.
In a ``shifted dipole" (SD) fluid, two opposite charges ($|Q|=0.2\,Q_{\mathrm{LJ}}$) lie on opposite ends of a line crossing the LJ center, at distances \mbox{$L_Q/\sigma = 0.125$} and \mbox{$L_{-Q}/\sigma = 0.375$} away from the center.
In a ``Stockmayer-like'' (STM) fluid, the same charges lie on a similar line, now at the same distance \mbox{$L/\sigma = 0.25$} from the center.
Similar to the FA fluid, the sites of the point charges in SD and STM do not have LJ interactions. 

We confine the five fluids between two planar and parallel walls constructed of LJ particles on a face-centered cubic lattice with a solid density of $\rho_{\mathrm{s}}\sigma_{\mathrm{s}}^{3}=1.07$, where we use $\sigma_{\mathrm{s}}=\sigma$.
For molecules with $\sigma_{\beta}=0$ and $\varepsilon_{\beta}=0$, we set the LJ parameters of solid--fluid interaction to zero.
Otherwise, the size parameter of the solid--fluid interaction is calculated with the Lorentz rule.
The interaction energies $\varepsilon_{\mathrm{sf},i}$ between the wall and atoms $i$ in a molecule are set proportional to the energy parameters of these atoms: 
we use $\varepsilon_{\mathrm{sf},i}=\zeta\varepsilon_{i}$, where $\zeta$ is the solid--fluid interaction parameter~\cite{becker_lm_2014,fertig_jpcc_2024,heier_jced_2018}.
The aforementioned study by Becker~\cite{becker_lm_2014} showed that, for LJ truncated-and-shifted droplets, the contact angle was $90^\circ$ when $\zeta=0.5$.
For weaker solid--fluid interactions ($\zeta<0.5$), the wall became solvophobic (contact angle $>90^\circ$), whereas, for stronger solid--fluid interactions ($\zeta>0.5$), the wall became solvophilic (contact angle $<90^\circ$).
As all five fluids have small off-center dipoles, we expect that varying $\zeta$ results in solid--fluid interfaces qualitatively similar to those found in \mbox{Becker \textit{et al}}.

\subsection{Simulation method and observables}

Using the August 2023 version of the LAMMPS simulation package \cite{LAMMPS}, we perform MD simulations of 4753 molecules of the five different fluids, between two walls, each five atomic layers wide.
We set the LJ interaction cutoff at $4\sigma$ and the long-range electrostatic interactions are handled with a PPPM $k$-space solver with a relative accuracy of $10^{-4}$.
We use the SHAKE algorithm~\cite{ryckaert_jcp_1977} to constrain bond lengths, yielding rigid molecules.
Additionally, the solid is rigid throughout the whole simulation.
The time step for the time integration is $\Delta t \sqrt{\varepsilon/m}/\sigma=4\times 10^{-4}$, and a Langevin thermostat controls the temperature of the fluid.
\begin{figure}
    \centering
	\includegraphics[width =\linewidth]{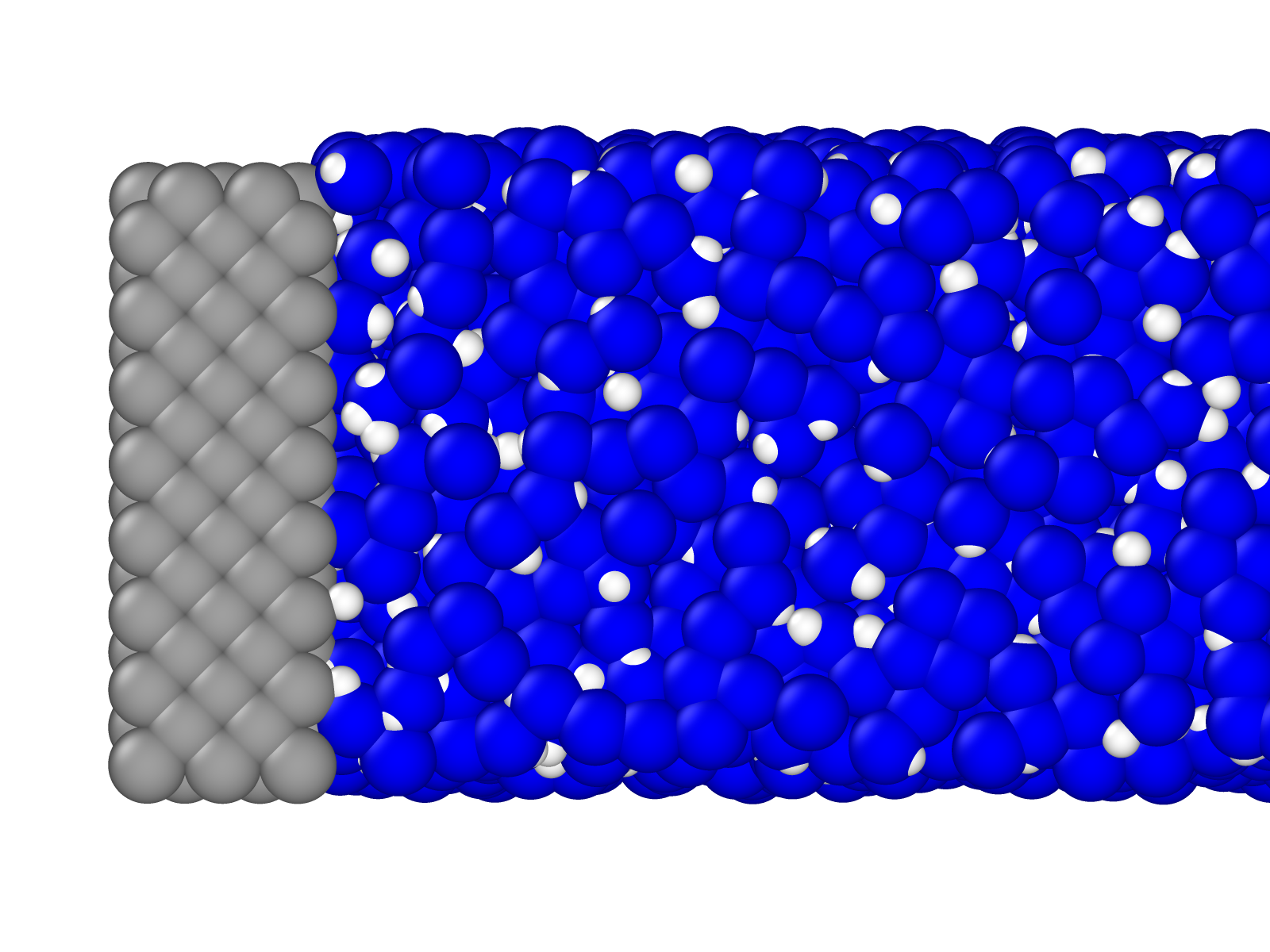}
    \caption{Half-cell of the simulation box. Grey, blue, and white spheres correspond to solid atoms, atoms $\alpha$ and $\beta$ of the FA molecules respectively. The atoms are not depicted to scale.}\label{fig:paralel}
\end{figure}
For subcritical temperatures, we set the wall's surface area $A$ to $A/\sigma^2$=154.16; for supercritical temperatures, we use $A/\sigma^2$=962.24.
We use a Cartesian coordinate system with the $x$ and $y$ coordinates lying in the plane of the solid wall, the $z$ coordinate running orthogonally between the walls, with the origin placed at the edge of the left solid---at a distance $0.5\sigma$ right of the center of the right-most atoms of the left wall.
Similar to Ref.~\cite{nickel_prl_2024}, we account for the slab geometry in the $z$-direction~\cite{yeh_jcp_1999}.
In the $x$- and $y$-direction, periodic boundary conditions are applied.

In the equilibration runs, we set the pressure, and thus the fluid density, by exerting a force on the atoms of the right plate corresponding to the same reduced pressure, $p/p_{\mathrm{c}}=1$ (see \cref{appendix:vle} for more details) for all five fluids.
We then let the system pressurize and equilibrate for $3-5\times10^6$ time steps.
This results in wall-wall separations of about $40-90\,\sigma$.
An exemplary, equilibrated set-up is shown on \cref{fig:paralel}.

After equilibration, we keep the wall--wall separation constant, resulting in an NVT simulation. 
The production runs last $6\times10^6$ time steps.
In the production run, area-averaged profiles are sampled in the $z$-direction normal to the wall surface; the molecule density $\rho$ and charge density $q$ is sampled in bins of $\Delta z/\sigma=0.03$ wide, from a plane at $z_0/\sigma=-1.6$ within the wall up to a plane at $24.4\sigma$ far in the fluid.
Both the molecular and charge densities are sampled as block-averages, in $10^6$ time steps.
The initial profiles are discarded, and the subsequent five profiles are averaged to reduce statistical uncertainties. 
Note that, since all shown profiles of the density, charge density, and potential are area-averaged quantities, we will from here on not explicitly refer to them as ``area-averaged''.

We calculate the excess density as $\rho_{\mathrm{ex}}(z)=\rho(z)-\rho_{\mathrm{bulk}}$, where we subtract the bulk density $\rho_{\mathrm{bulk}}$ from the sampled density profiles.
Neglecting the influence of lateral variations in the charge density, we obtain the electric potential profile from the charge density profiles with the double integrated form of Poisson's equation,
\begin{equation}
    \psi(z) = -\dfrac{1}{4\pi\epsilon_0}\int\limits_{z_{0}}^{z}\int\limits_{z_{0}}^{z^{\prime}} q(z^{\prime\prime})\diff z^{\prime\prime} \diff z^{\prime}.
    \label{Eq:P}
\end{equation}
The surface potential $\chi$ is then evaluated from the potential profiles in the bulk, where the charge oscillations are decayed.
Similar to Ref.~\cite{nickel_prl_2024}, \mbox{$\chi=\psi_{\mathrm{bulk}}-\psi(z_0)$}.
The reference potential $\psi(z_0)$ in the solid is set to zero.
The surface potential $\chi$ reflects the potential difference between bulk liquid and the solid.
We find below that even at low temperatures, oscillations decay within $10-12\,\sigma$.
Therefore, for all presented results, we obtain $\chi$ by averaging $\psi_{\mathrm{bulk}}$ from $z/\sigma=12$ to $24$.

When we compare the different fluids below, we ensure that we compare the states of two fluids at equal reduced temperature $T/T_{\mathrm{c}}$, where $T_{\mathrm{c}}$ is the critical temperature of the respective fluid.
We obtained the critical temperatures from MD simulations of the vapor--liquid equilibria of all the diatomic model fluids, see \cref{appendix:vle}. 
These yielded the critical temperatures \mbox{$T_{\mathrm{c}}^{\mathrm{FA}}k_B/\varepsilon=1.3$}, \mbox{$T_{\mathrm{c}}^{\mathrm{MA}}k_B/\varepsilon=1.8$}, and \mbox{$T_{\mathrm{c}}^{\mathrm{FS}}k_B/\varepsilon=2.55$} (and corresponding critical pressures and densities).
We assume that the critical properties of the STM and SD fluids are similar to those of the FA fluid, because of their small dipole moments and the similarity of the molecular geometries; therefore, we are not carrying out vapor--liquid equilibrium simulations for SD and STM.

\section{Results and discussion}
\subsection{Solvophobicity/solvophilicity of the walls}
\label{sec:solv}

\begin{figure}
    \centering
	\includegraphics[width =\linewidth]{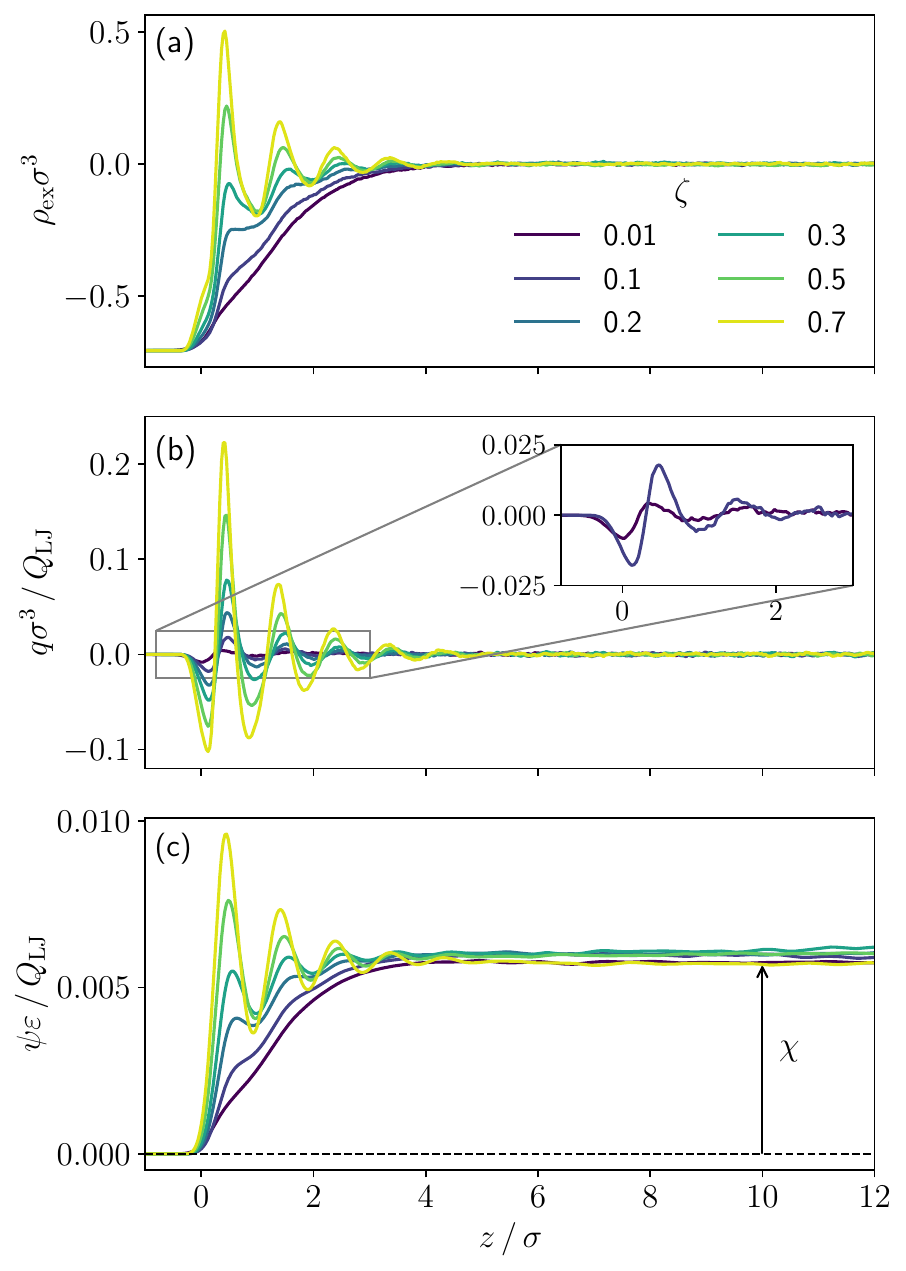}
    \caption{(a) Excess density $\rho_{\mathrm{ex}}$, (b) charge density $q$, and (c) electric potential $\psi$ profiles as a function of the distance $z$ from the solid--fluid interface, for various solid--fluid interaction parameter $\zeta$ values, for the FA fluid at ${T/T_{\mathrm{c}}=0.75}$. The surface potential $\chi$ is indicated with an arrow. The inset in (b) shows $q(z)$ results for $\zeta=0.01$ and $0.1$.}\label{fig:all_vs_zeta}
\end{figure}
First, we study how the solid's solvophilicity, characterized by the solid--fluid interaction parameter \mbox{$\zeta=\varepsilon_{\mathrm{sf}, i}/\varepsilon_i$}, affects the fluid's structure near the solid--fluid interface.
To this end, we examine in \Cref{fig:all_vs_zeta} results obtained for the FA fluid at a subcritical temperature (\mbox{${T/T_{\mathrm{c}}=0.75}$}) and various values of the solid--fluid interaction parameter~($\zeta=0.01-0.7$).

\Cref{fig:all_vs_zeta}(a) shows the excess density profile.
For \mbox{$\zeta\ge 0.3$}, the density profiles exhibit local maxima, with the height of the oscillation peaks increasing with increasing solid--fluid interaction parameter.
Given that atom $\beta$ is sizeless in the FA fluid, the maximum nearest the solid corresponds to atom $\alpha$ being ``locked in'' at a distance to the surface that corresponds approximately to its radius, $\sigma/2=0.5$.
For \mbox{$\zeta=0.01 - 0.2$}, fluid depletion zones are found near the interface.
These depletion zones indicate that interactions between fluid molecules are favored energetically over solid--fluid interactions.
In all of these cases (most clearly for $\zeta=0.2$), the depletion layer is superimposed by weak density oscillations, which indicates the onset of adsorption.
We note that the density profile has small nonzero values even for negative $z$.
This is because the $\beta$ atoms can penetrate the voids left between the spherical atoms in the outermost wall layer.

\Cref{fig:all_vs_zeta}(b) shows the charge density $q(z)$.
Damped charge oscillations are found for all considered values of~$\zeta$.
The inset of \Cref{fig:all_vs_zeta}(b) shows weak charge oscillations, even for the most solvophobic walls.
At larger distances from the wall, thermal motion restores a uniform fluid density and charge neutrality, which can be seen in the damping of the oscillations in both the density and charge profiles.
We find below similar charge oscillations for low-density, high-temperature systems.
The charge density peaks for \mbox{$\zeta\ge 0.2$} are located at the same $z$ values as the density peaks in \cref{fig:all_vs_zeta}(a).
The positive charge density peak reflects the localization of FA molecules in the first molecular layer, which, as discussed above, is determined by the radius of the $\alpha$ atoms.
Correspondingly, the negative values between the positive maxima arise from the $\beta$ atoms rotating around the ``locked in'' $\alpha$ atoms.
This means that the charge oscillations result from the structural asymmetry of the FA molecules.
This result is qualitatively consistent with the findings in Ref.~\cite{nickel_prl_2024} for the SPC/E water model.
This is not surprising as FA molecules resemble SPC/E water: both molecules contain interactionless LJ sites (atom $\beta$ for FA, hydrogen atoms for SPC/E water).

\Cref{fig:all_vs_zeta}(c) shows the electric potential profiles~$\psi(z)$.
The interfacial structure of the potential resembles the excess density profiles.
Moreover, the surface potential $\chi$ is almost the same for all considered solid--fluid interactions $\zeta$, even though these span almost two decades, encompassing both solvophilic and solvophobic walls.
This suggests that $\chi$ depends predominantly on fluid properties (addressed below), indicating why Ref.~\cite{nickel_prl_2024} found a similar surface potential at the water--graphite interface as Ref.~\cite{chapman_pccp_2022} found for the water vapor--liquid interface.
As the solid--fluid interaction parameter $\zeta$ hardly affects the surface potential, we fix it to $\zeta=0.5$ from here on.

\begin{figure}
    \centering
	\includegraphics[width =\linewidth]{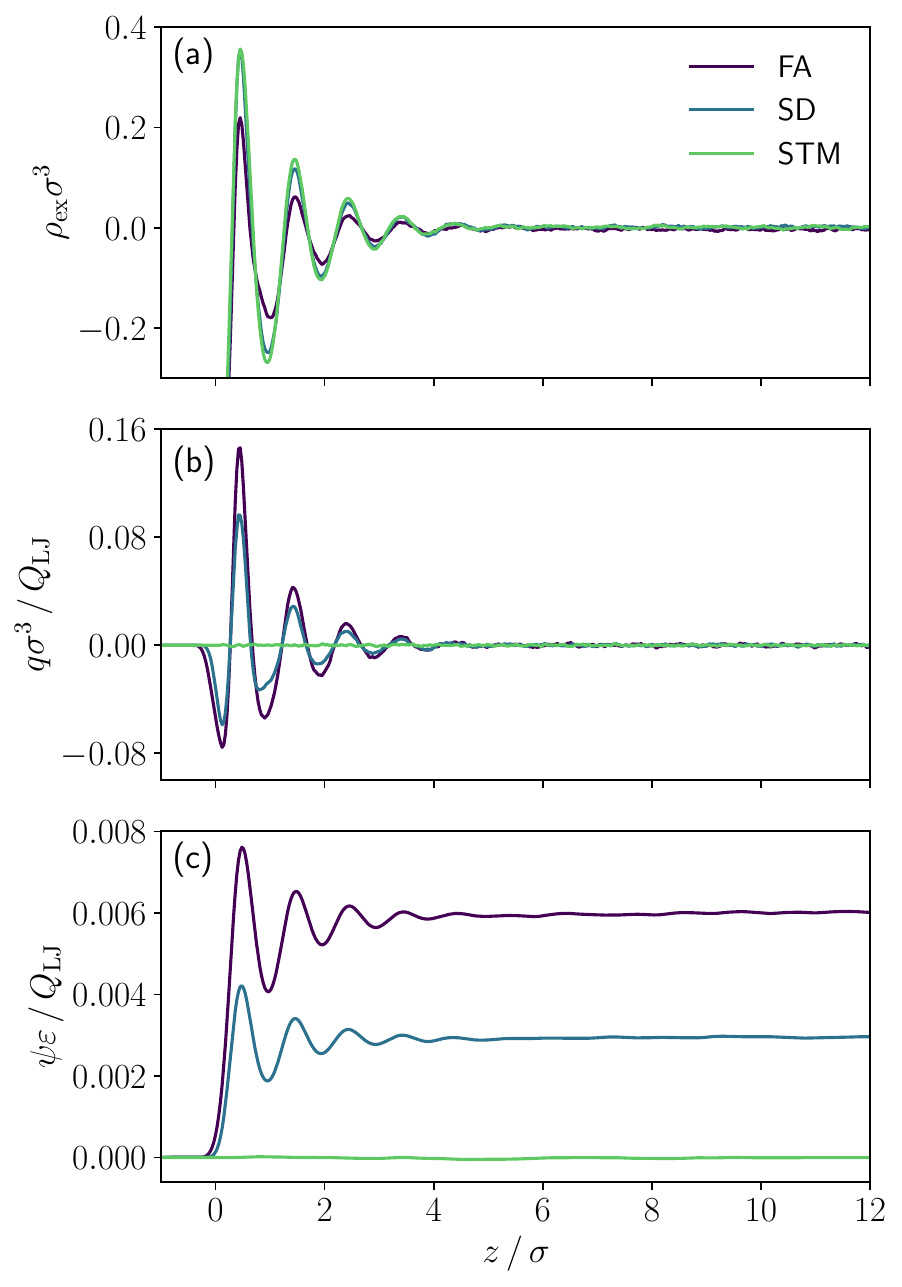}
    \caption{(a) Excess density $\rho_{\mathrm{ex}}$, (b) charge density $q$, and (c) electric potential $\psi$ profiles as a function of the distance~$z$ from the solid--fluid interface for the FA, SD, and STM fluids [see \cref{fig:particles}] at reduced temperature $T/T_{\mathrm{c}}=0.75$.}\label{fig:wheredip}
\end{figure}

\subsection{Location of the dipole moment}\label{section:stm}
Having inspected the influence of solid--fluid interactions on the surface potential for the FA fluid, we now turn to the three monoatomic fluids (FA, STM, and SD) and study the effect of their different charge placement relative to the LJ center.
The STM fluid has mirror-symmetric charge placement, while the partial charges in the SD and FA molecules are placed around a point that is offset with respect to the LJ center (see \cref{fig:particles}).

The excess density profiles of all three fluids [\cref{fig:wheredip}(a)] have a similar oscillatory structure.
The density oscillations are stronger for the SD and STM fluids, with only minor differences between the two, than for the FA fluid.
For all three fluids, these oscillations decay over a similar length scale, $\sim4\,\sigma$.
As all three fluids have identical LJ atoms, differences in $\rho_{\mathrm{ex}}$ must be caused by their different Coulomb interactions.
$\beta$ atoms of FA molecules are placed at the edge of these molecules, so the charges on these atoms can approach each other closer than the those of SD and STM molecules.
This leads to occasional strong repulsions and, ultimately, weaker adsorption of the FA fluid than of the other two fluids. 

The different placement of the charges results in quite different interfacial charge density structures.
\Cref{fig:wheredip}(b) shows that the FA and SD fluids exhibit charge density oscillations of a similar magnitude, whereas the STM fluid is charge-neutral everywhere.
The charge oscillations are more pronounced for the FA than for the SD fluid, despite the larger density variations for the SD fluid.
This trend (oscillations strongest for FA, absent for STM) can be attributed to the offsets of the center of charge from the Lennard-Jones center, with STM exhibiting zero offset and FA exhibiting the largest offset.
Accordingly, \cref{fig:wheredip}(c) shows that the surface potential is larger for molecules with a more asymmetric charge distribution.
This has been also found in Ref.~\cite{varner_arxiv_2025} for the vapor--liquid interface of Stockmayer and off-center dipole fluids (see Figs. 8 and 9 therein).

\begin{figure}
    \centering
	\includegraphics[width =\linewidth]{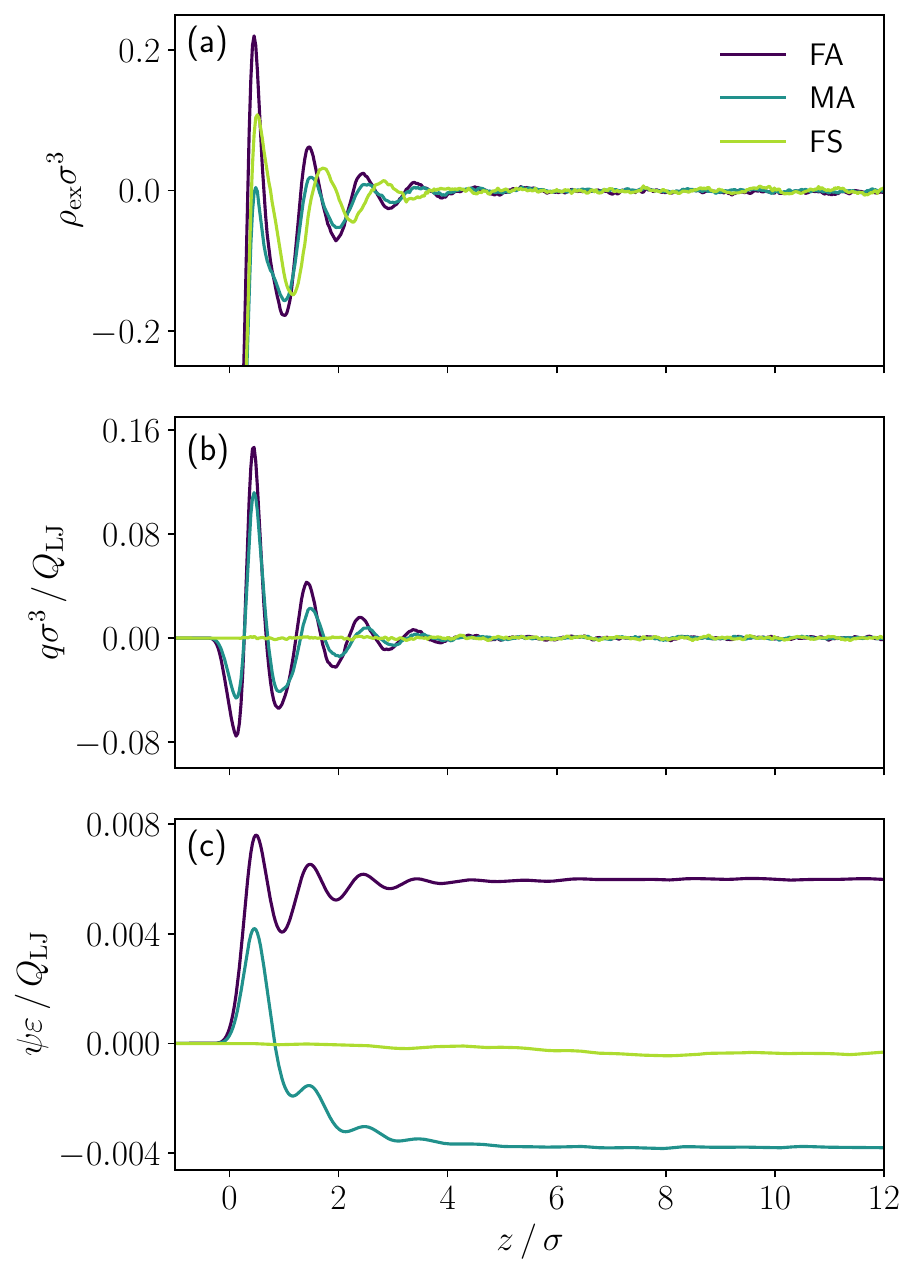}
    \caption{(a) Excess density $\rho_{\mathrm{ex}}$, (b) charge density $q$, and (c) electric potential $\psi$ profiles as a function of the distance $z$ from the solid--fluid interface for fluids FA, MA, and FS [see \cref{fig:particles}] for reduced temperature $T/T_{\mathrm{c}}=0.75$.}\label{fig:asymfl}
\end{figure}

\subsection{Molecular asymmetry}

Next, we compare the interfacial structure and surface potentials of the FA, MA, and FS fluids---all containing dipolar molecules, but with different molecular geometries, see \cref{fig:particles}.
\Cref{fig:asymfl}(a) shows that, at $T/T_{\mathrm{c}}=0.75$, these fluids have oscillatory interfacial excess density profiles similar to those of LJ fluids~\cite{toxvaerd_fscs_1981,becker_lm_2014,fertig_jpcc_2024}.
The length scale over which the density oscillations decay is similar for all three fluids.
The FA fluid exhibits the strongest oscillatory layering, followed by FS and then MA.
The density oscillations near the interface have smaller amplitude for the FS fluid than for the FA fluid; they are also broader and further apart. 
While the latter clearly reflects the larger size of the FS molecules, the former may be attributed to the smaller bulk density of the FS fluid at the same reduced temperature.
For the MA fluid, we do not find a positive excess density in the first adsorption layer.
Nevertheless, there is a pronounced local maximum at $z/\sigma\approx 0.5$.
This indicates that, despite the finite radius of the $\beta$ atom, localization in the first molecular layer of the MA fluid is still predominantly determined by $\sigma_{\alpha}$.
The density profile also exhibits a shoulder at $z/\sigma\approx 0.8$, however, which can be attributed to configurations in which the $\beta$ atom points towards the wall.

Next, \cref{fig:asymfl}(b) shows the charge density profiles of the FA, MA, and FS fluids, again at $T/T_{\mathrm{c}}=0.75$.
(We showed and discussed the charge oscillations for the FA fluid in \cref{sec:solv}.)
The FA fluid exhibits the strongest charge oscillations, the MA fluid shows slightly weaker oscillations, and the FS fluid does not exhibit significant oscillations.
The charge density profiles of the FA and MA fluids decay on a similar length scale.
For the MA fluid, there is a pronounced positive peak in the charge density \mbox{at $z/\sigma\approx 0.5$}. Like discussed above for the FA fluid, this can be attributed to $\alpha$ atoms being localized in the first molecular layer near the interface.
For the MA fluid, however, the negative charge density minima to the left and right of the first positive peak are almost at level, and the right minimum is substantially broader than the left.
This difference of the MA fluid compared to the FA fluid can be attributed to steric hindrances making an orientation of the $\beta$ atom towards the wall unfavorable.
Because of this steric hindrance, the charge oscillation becomes weaker at the interface, thus also altering the surface potential $\chi$ (see below).
Furthermore, the Coulombic repulsion between $\beta$ atoms of two MA molecules is on average weaker than that between $\beta$ atoms of FA molecules.
Hence, orientational ordering near the solid--fluid interface is weaker for the MA fluid than for the FA fluid.
The absence of charge density variations for the FS fluid is due to the equal LJ interaction parameters of the two atoms.
Since both atoms interact equally with the solid, there is no preference for one of them; hence, there are no local charge imbalances, and no surface potential [see \cref{fig:asymfl}(c)].
These findings for the FS model fluid correspond closely to the findings of \cref{section:stm} for the STM fluid: when there is no preferential adsorption of one of the fluid's equal-size atoms to the wall, there will be no charge oscillations.

\begin{figure}
    \centering
	\includegraphics[width =0.5\linewidth]{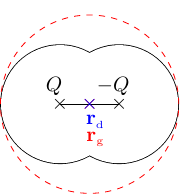}\\
    \includegraphics[width =0.32\linewidth]{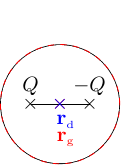}
    \includegraphics[width =0.4\linewidth]{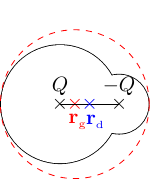}
    \caption{Schematics for FS (top), STM (bottom left), and MA (bottom right) molecules. The dashed, red circle shows the sphere with the smallest radius that envelopes the molecule. We take the center of this sphere as the geometric center of the molecule $\bf{r}_{\mathrm{g}}$, and $\bf{r}_{\mathrm{d}}$ is the dipolar center.}\label{fig:geom}
\end{figure}

\begin{figure*}[t]
    \centering
	\includegraphics[width =\linewidth]{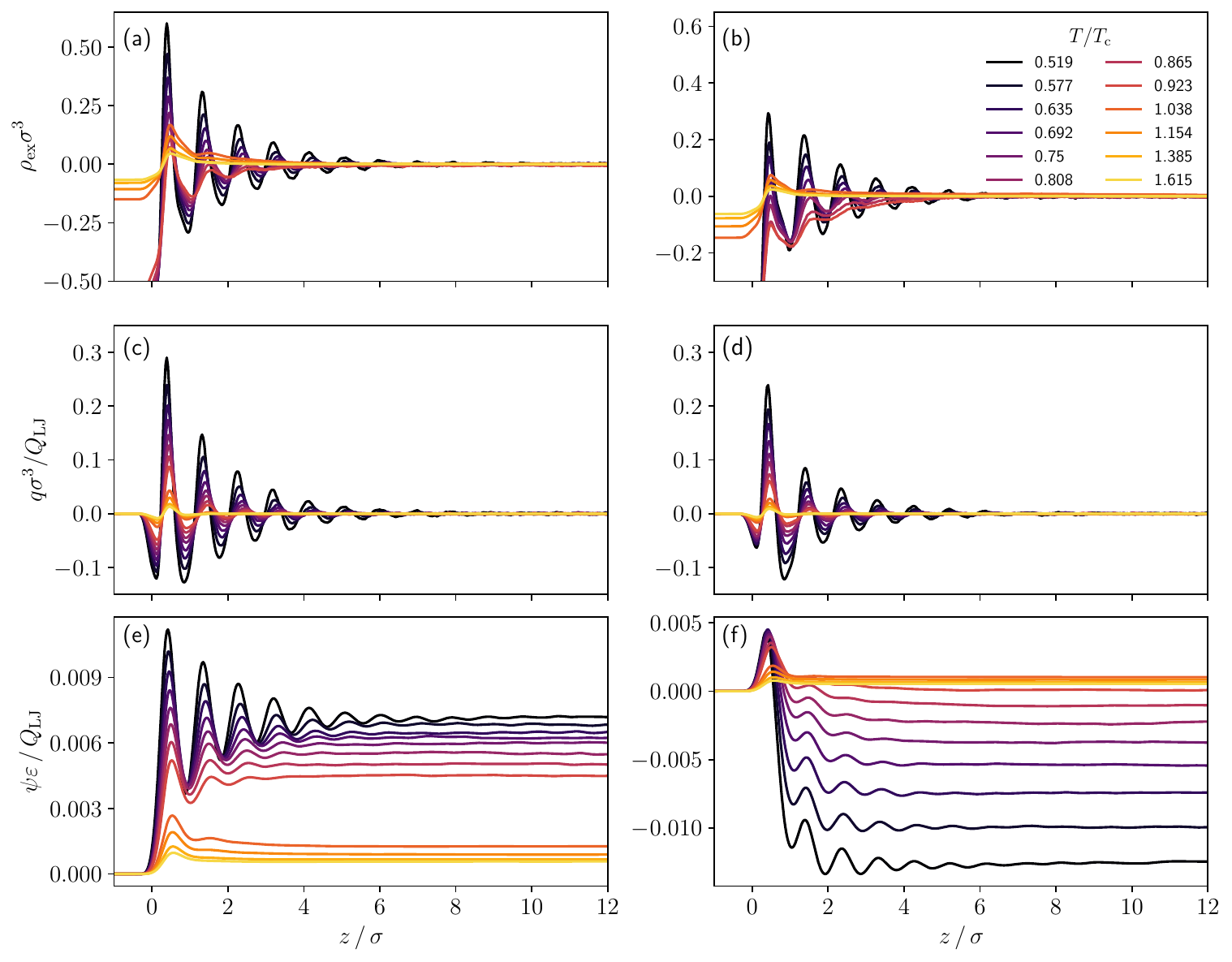}
    \caption{Excess density $\rho_{\mathrm{ex}}$ (top row), charge density $q$ (middle row), and electric potential $\psi$ (bottom row) for fluids FA (left column) and MA (right column) as a function of the distance $z$ from the solid--fluid interface, for various reduced temperatures in the range $T/T_{\mathrm{c}}=0.519-1.615$.}\label{fig:Tdepprof}
\end{figure*}

The symmetry considerations made in our discussion of~\cref{fig:wheredip} and~\cref{fig:asymfl} can be brought to a common denominator by comparing the positions of two characteristic points in the molecules, as illustrated in \cref{fig:geom} for the FS, STM, and MA fluids. 
The first of these points is the molecule's geometric center $\bf{r}_{\mathrm{g}}$, which we define as the center of the smallest possible enveloping sphere (dashed red line). 
The second point is the dipolar center $\bf{r}_{\mathrm{d}}$, \textit{i.e.}, the point halfway between the two opposite partial charges.
Based on our observations in \cref{fig:wheredip}(b) and \cref{fig:asymfl}(b), we hypothesize that there will generally be no charge oscillations, and thus no surface potential, if the geometric and dipolar centers align.
Vice versa, we hypothesize that an off-center dipole moment (\mbox{$\bf{r}_{\mathrm{d}}\neq \bf{r}_{\mathrm{g}}$}) is required for non-zero surface potentials to appear.
Furthermore, we hypothesize that any dipolar molecular liquid with an off-center dipole---as present in virtually all real polar molecules---would result in interfacial charge oscillations, and non-zero surface potentials.
We expect that our considerations can be extended to molecules with more atoms and/or more complex charge distributions, by considering a (weighted) average of the partial charge locations.

\subsection{Effect of temperature on the fluid structure}
In the previous sections, we identified the origins of charge oscillations at solid--fluid interfaces. The cases considered so far all involved a dense molecular liquid.
In the following, we discuss the effect of temperature on the adsorption layer, charge density profile, and the surface potential $\chi$,
with emphasis on the influence of steric hindrances.
We focus, therefore, on the diatomic molecules FA and MA (the FS fluid does not produce a finite surface potential).
We vary the temperature in the range $T/T_{\mathrm{c}}=0.519-1.615$, from subcritical temperatures ($T<T_{\mathrm{c}}$) close to the freezing point of the fluids up to the supercritical regime ($T>T_{\mathrm{c}}$).
We keep the same pressure at all temperatures, so the following results are elements of an isobar. 
This means that, in general, the bulk density of the fluids decreases with increasing temperature, see \cref{appendix:rho} for bulk density as a function of temperature. 

\Cref{fig:Tdepprof} reports the excess density, charge density, and potential profiles for various temperatures, for FA (left column) and MA (right column).
\Cref{fig:Tdepprof}(a) and~(b) show that oscillations in the excess density profiles become weaker with increasing temperature, which can be attributed to the decreasing density along the isobar.
For both FA and MA, this weakening of the interfacial layering progresses gradually at the subcritical temperatures, whereas a rapid change in the interfacial structure occurs around $T_\mathrm{c}$:
for $T/T_\mathrm{c}=0.923$, four molecular layers are discernible from the density oscillations, whereas at $T/T_\mathrm{c}=1.038$, only the first molecular layer and a weak second peak can be recognized.
Furthermore, the extended fluid depletion zone found at $T/T_\mathrm{c}=0.923$ is no longer present in the supercritical regime.
These observations correspond to typical adsorption behavior in a low-density fluid.
At high temperatures, the thermal motion ``smears out" the density profiles.

\Cref{fig:Tdepprof}(c) and (d) show that, similar to the excess density profiles, the oscillation amplitude of both fluid's charge densities decreases with increasing temperature. 
Like in the previous section [see \Cref{fig:asymfl}(b)], we find larger charge oscillation amplitudes for the FA fluid than for MA.
An important difference between the two fluids is the shallower first charge dip for the MA fluid.
This can be explained with the differences of $\sigma_{\beta}$:
for the FA fluid, the $\beta$ atom can penetrate the solid, whereas for the MA fluid, rotation is sterically hindered.

\begin{figure}[t]
    \centering
	\includegraphics[width =\linewidth]{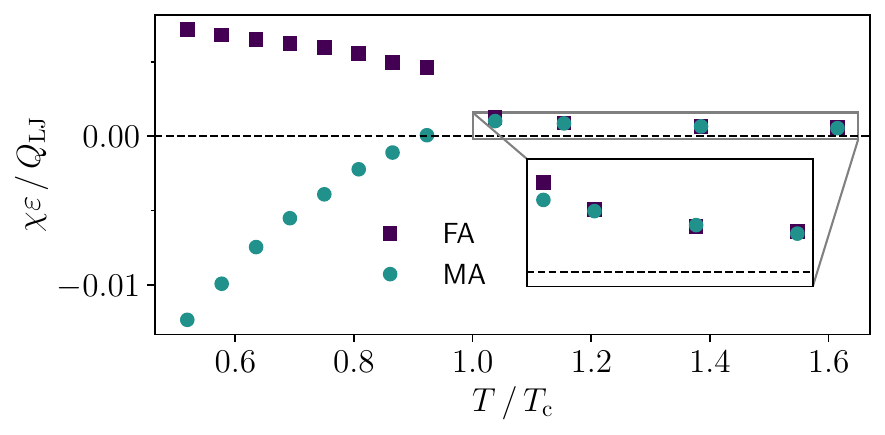}
    \caption{Surface potential $\chi$ as a function of reduced temperature $T/T_{\mathrm{c}}$ for fluids FA and MA [\cref{fig:particles}]. The dashed line indicates $\chi=0$. The inset shows the data points for supercritical temperatures.}\label{fig:chi_vs_T}
\end{figure}

A comparison of the potential profiles for the FA and MA fluids in \cref{fig:Tdepprof}(e) and (f) reveals that, for both fluids, the oscillations weaken with increasing temperature.
From the profiles in \cref{fig:Tdepprof}(e) and (f), we determine the temperature dependent surface potential $\chi(T)$, which we plot in \cref{fig:chi_vs_T}.
For the FA fluid, we observe that \mbox{$\diff\chi^{\mathrm{FA}}/\diff T< 0$} for all temperatures.
For the MA fluid, in turn, we see that \mbox{$\diff\chi^{\mathrm{MA}}/\diff T> 0$} for subcritical temperatures, and \mbox{$\diff\chi^{\mathrm{MA}}/\diff T<0$} for supercritical temperatures.
We did not carry out simulations at temperatures close to the critical temperature, but we expect decreasing $\chi$ for the FA fluid and a sign change of $\diff\chi^{\mathrm{MA}}/\diff T$.
At supercritical temperatures, we obtain similar values of $\chi$ for the FA and the MA fluid, as well as similar $\diff\chi/\diff T$.
In this regime, the MA molecules can rotate almost freely; hence, the MA fluid starts behaving like the FA fluid, resulting in similar surface potentials.

To connect \cref{fig:chi_vs_T} to physical units, we use as an example the LJ parameters of SPC/E water ($\varepsilon/k_{\mathrm{B}}=\SI{78.2}{\kelvin}$ and $\sigma=\SI{3.166}{\angstrom}$).
From this, we obtain the LJ unit of charge, \mbox{$Q_{\mathrm{LJ}}^\mathrm{SPC/E}\approx\SI{6.17e-21}{\coulomb}$}, as well as the LJ unit of electric potential, $\psi_{\mathrm{LJ}}^\mathrm{SPC/E}=\SI{0.175}{\volt}$.
In these terms, the partial charges on the sites of the model molecules are $|Q|=0.2Q_{\mathrm{LJ}}\approx 0.01\, e$, where $e$ is the elementary charge.
The obtained potentials ($\chi\sim 0.01\,\psi_\mathrm{LJ}$), in turn, correspond to potential jumps of the order of millivolts.
In SPC/E water, the value of the partial charges are $q_{\mathrm{O}}=-0.8476\, e$ and $q_{\mathrm{H}}=0.4238\, e$, \textit{i.e.}, roughly $50$ and $100$ times larger than the charges used here.
If, for a simple comparison, we assume that $\chi$ scales linearly with the partial charges, we reach the order of hundred millivolts, which is the right order of magnitude\cite{nickel_prl_2024, chapman_pccp_2022}.
Note that this linear scaling may not be accurate, as increasingly stronger Coulomb interactions can lead to a nonlinear increase in~$\chi$.

\section{Conclusion}
We analyzed the solid--fluid interfacial structure for different model fluids confined between two slabs of a model solid, and identified microscopic requirements for an interfacial charge oscillation at the solid--fluid interface and a non-zero surface potential.
We argued that, in dipolar fluids, the molecules' dipolar center must differ from their geometric center to create charge oscillations and a non-zero surface potential difference across the interface. 
If this hypothesis is true, then studies of systems with a centered dipole moment, such as the Stockmayer fluid or the dipolar hard sphere fluid, are not suitable to describe the interfacial charge ordering of real polar fluids, such as water.

Furthermore, we found that variations in the solid--fluid interaction parameter had no distinct effect on the surface potential $\chi$, despite a very clear influence on the interfacial structure.
Additionally, we found that, at subcritical temperatures, the surface potentials of different model fluids, as well as their temperature dependence, can differ strongly, whereas they coincide at supercritical temperatures. 
We found that the absolute value of the surface potential decreases with increasing temperature or the fluid's density.

Experimental observation of charge oscillations at the solid--fluid interface is not possible yet.
Model systems such as those considered here may help to understand the solid--liquid interfacial structure of, for example, aqueous systems.
While we studied confined simple model fluids, future work could consider fluids and walls with hard interactions, which would provide insights at an even simpler level.

 \newpage
 \section{Acknowledgements}
All authors were supported by a FRIPRO grant from The Research Council of Norway (Project No. 345079).
The authors acknowledge the Orion High Performance Computing Center (OHPCC) at the Norwegian University of Life Sciences (NMBU) for providing computational resources that have contributed to the research results reported within this paper.

\section{Data availability}
All LAMMPS scripts and the generated data are available at (link)
\appendix
\renewcommand\thefigure{\thesection\arabic{figure}} \setcounter{figure}{0}

\section{Vapor-liquid equilibrium calculations and critical parameters}\label{appendix:vle}
\begin{figure}
    \centering
	\includegraphics[width =\linewidth]{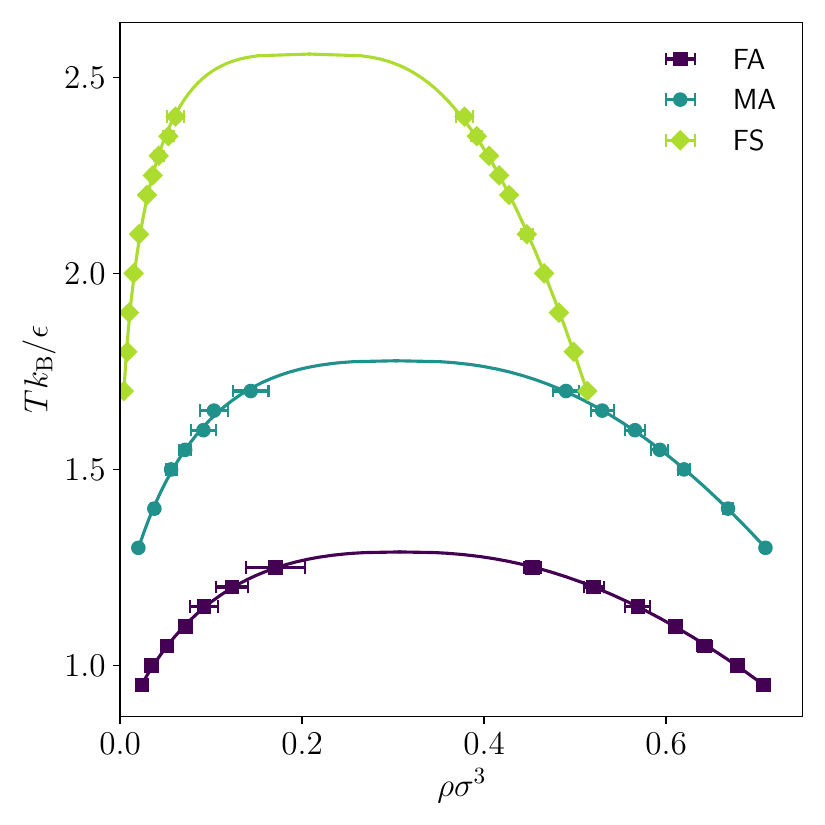}
    \caption{$\rho-T$ projection of the vapor-liquid phase equilibrium of fluids FA, MA, and FS as obtained by MD (symbols). The lines are fits to these data using \cref{eq:fit}.}\label{fig:vle}
\end{figure}
We performed direct vapor--liquid equilibrium calculations for fluids FA, MA, and FS using MD simulations.
The simulation cell was a rectangular box with dimensions $15\,\sigma\times15\,\sigma\times100\,\sigma$, divided into three regions.
The middle region contained a fluid at density $\rho_{\mathrm{L}}$ and pressure $p_{\mathrm{L}}$ and the surrounding regions a fluid at lower density $\rho_{\mathrm{V}}$ and pressure $p_{\mathrm{V}}$.
As we were not aware of equations of state for these three fluids, 
we guessed initial values for $\rho_{\mathrm{L}}$ and $\rho_{\mathrm{V}}$.
The same thermostat, cut-off lengths, and other system parameters were used as in the main body of the article.

The system was equilibrated for $5\times10^6$ time steps, followed by a production run of $5\times10^6$ time steps.
In the production run, the densities and pressures of the vapor and liquid phase were sampled in rectangular boxes.
Time averages of the molecular density and the pressure were gathered for $5\times10^5$ time steps, resulting in 10 values for $\rho_{\mathrm{V}}$, $\rho_{\mathrm{L}}$, $p_{\mathrm{V}}$, and $p_{\mathrm{L}}$.
The averages and the uncertainties of the vapor--liquid equilibrium properties are reported on \cref{fig:vle} for the three fluids.
The uncertainties of $\rho_{\mathrm{V}}$ increase with increasing temperature, especially in the vicinity of the critical point. 

The critical temperatures $T_{\mathrm{c}}$ were then evaluated by fitting the MD data with
\begin{align}\label{eq:fit}
    \rho_{\mathrm{L}}-\rho_{\mathrm{V}}=B\left(1-\dfrac{T}{T_{\mathrm{c}}}\right)^{\beta}
\end{align}
where $B$, $\beta$, and $T_{\mathrm{c}}$ are fitting parameters with $\beta$ being the critical exponent.
The critical pressure $p_c$ can be evaluated through fitting the vapor pressure curve with, \textit{e.g.}, a quadratic function and evaluating the fitted function at the critical temperature. 
The critical densities $\rho_{\mathrm{c}}$ were evaluated also via fitting, with
\begin{align}\label{eq:fit2}
    \dfrac{\rho_{\mathrm{L}}+\rho_{\mathrm{V}}}{2}=\rho_{\mathrm{c}}+A(T_{\mathrm{c}}-T)
\end{align}
where $A$ and $\rho_{\mathrm{c}}$ are fitting parameters.

We obtained the following values: $T^{\mathrm{FA}}_{c}=1.2895\pm0.0217$, 
$p^{\mathrm{FA}}_{c}=0.1274\pm0.0035$, $\rho^{\mathrm{FA}}=0.3072\pm0.0066$ $T^{\mathrm{MA}}_{c}=1.7772\pm0.0359$, 
$p^{\mathrm{MA}}_{c}=0.1719\pm0.0038$, $\rho^{\mathrm{MA}}=0.3035\pm0.0056$, $T^{\mathrm{FS}}_{c}=2.5596\pm0.0577$, 
$p^{\mathrm{FS}}_{c}=0.1334\pm0.0034$, and $\rho^{\mathrm{FS}}=0.2079\pm0.0022$.
In the main body of the study we used $T^{\mathrm{FA}}_{c}=1.3$, $T^{\mathrm{MA}}_{c}=1.8$, and $T^{\mathrm{FS}}_{c}=2.55$, and $p^{\mathrm{FA}}_{c}=0.13$, $p^{\mathrm{MA}}_{c}=0.17$, and ${p^{\mathrm{FS}}_{c}=0.13}$.
The critical parameters of the FA fluid are close to those of the Lennard-Jones fluid~\cite{stephan_jcim_2019}.
This is not surprising, as the FA fluid is a LJ fluid with a small, off-center dipole.
\renewcommand\thefigure{\thesection\arabic{figure}} \setcounter{figure}{0}

\section{Temperature dependent bulk densities}\label{appendix:rho}
\begin{figure}[htb]
    \centering
	\includegraphics[width =\linewidth]{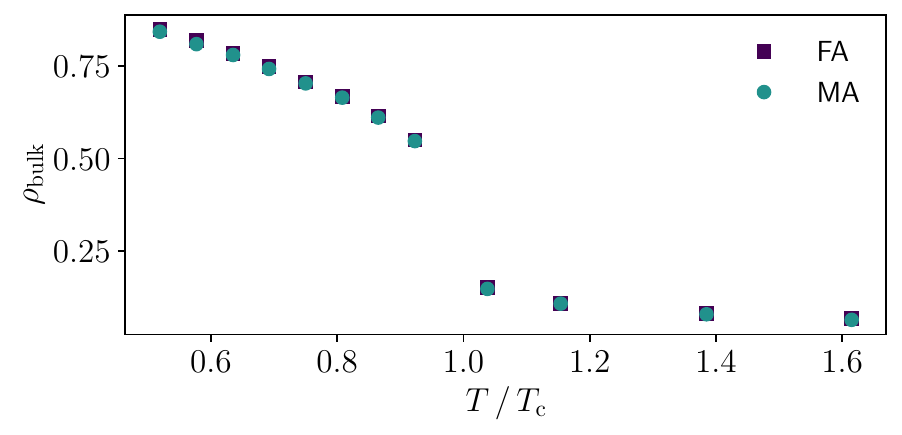}
    \caption{Bulk density $\rho_{\mathrm{bulk}}$ as a function of reduced temperature $T/T_{\mathrm{c}}$ at $p/p_{\mathrm{c}}=1$ for fluids FA and MA [\cref{fig:particles}].}\label{fig:rho}
\end{figure}
As expected from the law of corresponding states, and given that the critical densities of the two fluids are approximately the same (\mbox{$\rho^{\mathrm{FA}}_{\mathrm{c}}\approx\rho^{\mathrm{MA}}_{\mathrm{c}}$}), we find that the bulk densities of the two fluids are very similar in value for all investigated reduced temperatures.
Furthermore, the bulk densities decrease as temperature increases.
\bibliography{bibliography}
\bibliographystyle{unsrt}
\end{document}